\documentclass[prd,twocolumn,showpacs,floats,letterpaper,nofootinbib]{revtex4}
\usepackage{amsmath}
\usepackage{amssymb,latexsym,mathrsfs}
\usepackage{graphicx,subfig}

\usepackage{epsfig}
\usepackage{varioref,xr-hyper}
\usepackage{color}

\usepackage[hypertex,draft=false,verbose=false,debug=false
            hyperindex=true,hypertexnames=true]{hyperref}
\def\lsim{\;\raise 0.4ex\hbox{$<$}\kern -0.8em\lower 0.62 ex\hbox{$\sim$}\;}
\def\gsim{\;\raise 0.4ex\hbox{$>$}\kern -0.7em\lower 0.62 ex\hbox{$\sim$}\;}

\newcommand{\be}{\begin{equation}}
\newcommand{\ee}{\end{equation}}
\newcommand{\bs}{\boldsymbol}

\begin{document}

\title{Non-Gaussianity in WMAP Data Due to the Correlation of CMB Lensing Potential with Secondary Anisotropies}

\author{Erminia Calabrese$^{1,2}$, Joseph Smidt$^2$, Alexandre Amblard$^2$, Asantha Cooray$^{2}$, Alessandro Melchiorri$^1$,
Paolo Serra$^{2}$, Alan Heavens$^{3}$, Dipak Munshi$^{3,4}$}

\affiliation{$^1$Center for Cosmology, Dept. of Physics \& Astronomy, University of California Irvine, Irvine, CA 92697.}
\affiliation{$^2$Physics Department and INFN, Universita' di Roma ``La Sapienza'',
  Ple Aldo Moro 2, 00185, Rome, Italy}
\affiliation{$^3$Scottish Universities Physics Alliance (SUPA),~ Institute for Astronomy, University of Edinburgh, Blackford Hill,  Edinburgh EH9 3HJ, UK}
\affiliation{$^4$School of Physics and Astronomy, Cardiff University, CF24 3AA}

\begin{abstract}
We measure the skewness power spectrum of the Cosmic Microwave Background (CMB) anisotropies optimized for a detection of the secondary bispectrum generated by the correlation of the CMB lensing potential with integrated Sachs-Wolfe effect and the Sunyaev-Zel'dovich effect. The covariance
of our measurements is generated by Monte-Carlo simulations of Gaussian CMB fields with noise properties consistent with Wilkinson Microwave Anisotropy Probe (WMAP) 5-year data. When interpreting multi-frequency measurements we also take into
account the confusion resulting from unresolved radio point sources. We analyze Q, V and W-band WMAP 5-year raw and foreground-cleaned 
maps using the KQ75 mask out to $l_{\rm max}=600$. We find no significant evidence for a non-zero
non-Gaussian signal from the lensing-secondary correlation in all three bands and we constrain the overall amplitude
of the cross power spectrum between CMB lensing potential and the sum of SZ and ISW fluctuations to be 
0.42 $\pm$ 0.86 and 1.19 $\pm$ 0.86 in
combined V and W-band raw and foreground-cleaned maps provided by the WMAP team, respectively.
The point source amplitude at the bispectrum level measured with this skewness power spectrum is higher than 
previous measurements of point source non-Gaussianity. We also consider an analysis where we also account
for the primordial non-Gaussianity in addition to lensing-secondary bispectrum and point sources.
The focus of this paper is on secondary anisotropies.  Consequently the estimator is not optimised 
for primordial non-Gaussianity and  the limit we find on local non-Gaussianity from the foreground-cleaned V+W maps 
is $f_{\rm NL} = -13\pm 62$, when marginalized over point sources and lensing-ISW/SZ constributions to the total bispectrum.

\end{abstract}


\maketitle

\section{Introduction}

The all-sky, multi-frequency WMAP maps of Cosmic Microwave Background (CMB) anisotropies \cite{wmap5} have presented cosmologists with an
opportunity to test the cosmological scenario of structure formation at an unprecedented accuracy.
The results on the CMB temperature and polarization angular power spectra are in very good agreement with the
expectations of the standard cosmological model of structure formation based on primordial, adiabatic and scale
invariant perturbations, and cold dark matter \cite{Komatsu}. 

In addition to measuring the angular power spectrum and cosmological
parameter estimates from it \cite{Dunkley}, WMAP CMB maps are now routinely used to constrain statistical properties of the CMB
beyond the simple two-point angular correlation function. These studies include tests of cosmological isotropy \cite{Huterer,Eriksen,Hanson},
topology \cite{Luminet,Roukema}, and non-Gaussianity \cite{Smidt,Senatore,Pietrobon,Smith}, among other tests.
In the standard cosmological model, primordial CMB anisotropies are supposed to be Gaussian, however
non-Gaussianities may be present in the observed CMB maps through a combination of primordial non-Gaussianity of density perturbations
generated during inflation \cite{Maldacena:2002vr,Acquaviva:2002ud,Sasaki:1995aw,Lyth:2004gb,Lyth:2005fi}, the imprint of non-linear growth
of structures as probed by secondary effects \cite{Cooray1,Verde}, 
and mode-coupling effects by secondary sources of temperature fluctuations such as gravitational lensing of the CMB \cite{Cooray2,Goldberg}.

The detection of these non-Gaussian features will not only provide additional useful information on the
parameters of the standard cosmological model but also allow independent tests on constraining the amplitude of primordial
non-Gaussianity due to non-standard initial conditions and, ultimately, inflation after accounting for secondary non-Gaussian signals
generated since last scattering. Several recent studies have made use of the bispectrum for the study of non-Gaussianities \cite{Smith,Smidt}.
This quantity involves a three-point correlation function in Fourier or multipole space. The configuration
dependence of the bispectrum $B(k_1, k_2, k_3)$ with lengths ($k_1$, $k_2$, $k_3$) that
form a triangle in Fourier space can be used to separate various mechanisms for non-Gaussianities, depending on the effectiveness of the estimator used.

In most CMB non-Gaussianity studies \cite{wmap5,Yadav,Smith} the estimator employed involves a measurement that compresses all information of the bispectrum to a
single number called the cross-skewness computed with two weighted maps. Such a drastic compression limits the ability to study the angular dependence of the
non-Gaussian signal and to separate any confusing foregrounds from the primordial non-Gaussianity.
More recently, some of us have introduced a new estimator that preserves some angular dependence of
the bispectrum \cite{Cooray3,Munshi1}. This recently led to a new measurement of the primordial non-Gaussianity parameter \cite{Smidt}.

The skewness power spectrum is indeed a weighted statistic that can be tuned to study a particular
form of non-Gaussianity, such as what may arise either in the early Universe during inflation or late-times during structure formation, while retaining information
on the nature of non-Gaussianity more than the skewness alone.  When applied to the CMB data, this allows a way to explore all non-Gaussian signals,
including those generated  by contaminants such as Galactic foregrounds and unresolved point sources.

In this paper we analyze the recent WMAP data for the skewness power spectrum associated with cross-correlation between
lensing and ISW and SZ effects, respectively. The presence of a measurable signal in this secondary non-Gaussianity,
especially with the cross-correlation of lensing with the SZ effect, was identified in 2003 by Goldberg \& Spergel \cite{Goldberg,Cooray2}.
We provide first constraints on this signal from WMAP data using Q, V and W-band maps both in the ``Raw'' and foreground ``Clean'' form
as provided by the WMAP team publicly.

After accounting for the confusion from point sources generated by the overlap of the point source shot-noise bispectrum and the
lensing-secondary anisotropy cross-correlation bispectrum, we find no significant detection of the lensing effect in existing WMAP data.
We constrain the overall normalization of the lensing-SZ and lensing-ISW angular cross-power spectra to be 
0.42 $\pm$ 0.86 and 1.19 $\pm$ 0.86 in combined V and W-band raw and foreground-cleaned maps provided by the WMAP team, respectively.
The point source amplitude we determine from the raw map of Q-band with $b_{\rm src}=(67.8 \pm 5.4)\times 10^{-25}$ sr$^2$  is higher than
the estimate by the WMAP team with $(6.0 \pm 1.3) \times 10^{-5}$ $\mu$K$^3$-sr$^2$ \cite{Komatsu} (the value we determine is $(13.7 \pm 1.1) \times 10^{-5}$
$\mu$K$^3$-sr$^2$ in the same units used by the WMAP team). We find similarly a factor of 2 increase in the results from the V-band map.
In the case of clean maps, we find $b_{\rm src}=(6.2 \pm 5.4)\times 10^{-25}$ sr$^2$, which is smaller than the
WMAP team's estimate with clean maps for the Q band with  
$(4.3 \pm 1.3) \times 10^{-5}$ $\mu$K$^3$-sr$^2$ \cite{Komatsu} (the value we determine is $(1.4 \pm 1.1) \times 10^{-5}$
$\mu$K$^3$-sr$^2$ in the same units used by the WMAP team). We find similar differences in the V and W bands as well.
It is unclear exactly where these differences come from. We do not employ the same E-statistic that is optimized for point
sources as the WMAP team in the present study. 

We also considered the extent to which primordial non-Gaussianity confuse 
the detection of lensing-secondary correlations and found that when including $f_{\rm NL}$ in model fits,
in addition to point sources, leads to a factor of 2 degradation in the error of the amplitude of lensing-secondary correlation
power spectrum.

This paper is organized as follows: in the next Section, we review the measurement theory. We refer the reader to Munshi et al. \cite{Munshi2}
for more details. In Section~III we present our results and discuss the evidence for the secondary non-Gaussianity in WMAP data.

\section{Skewness Power spectrum Estimator}

If we consider three statistically isotropic fluctuation fields, say temperature anisotropies but weighted with different window functions differently,
 $X(\hat{\Omega})$, $Y(\hat{\Omega})$ and $Z(\hat{\Omega})$ described by the multipole moments $a^X_{l_1m_1}, a^Y_{l_2m_2},a^Z_{l_3m_3}$, all 
the information available in the three-point correlation function is contained in the angular bispectrum $B^{XYZ}_{l_1 l_2 l_3}$ defined by a triangle in multipole space 
with lengths of sides $(l_1,l_2,l_3)$ :
\begin{eqnarray}
B^{XYZ}_{l_1 l_2 l_3} = \sum_{m_1, m_2, m_3}
\left( \begin{array}{ccc}
l_1 & l_2 & l_3 \\
m_1 & m_2 & m_3 \\
\end{array} \right)
\langle a^X_{l_1m_1}a^Y_{l_2m_2}a^Z_{l_3m_3} \rangle \ .
\end{eqnarray}

Since measuring the full bispectrum is challenging, many previous measurements have focused mostly on the skewness which is collapse of information in the bispectrum
in some way to a single number. As discussed in Munshi et al.~\cite{Munshi2}, it is useful to pursue instead the skewness power spectrum which can be considered as the angular power
spectrum of the  correlation of the product map $X(\hat{\Omega})Y(\hat{\Omega})$ and the $Z(\hat{\Omega})$.
In the absence of sky-cut and instrumental noise, we can write the skewness power spectrum as :
\be
\langle a^{XY}_{lm} a^{Z}_{l'm'} \rangle \equiv C^{XY,Z}_l \delta_{l l'} \delta_{m m'} \ ,
\ee
where $a^{XY}_{lm}$ is the spherical harmonic transform coefficient of the field $XY$.

It is possible to show that this quantity, in the homogeneity and isotropy assumption, is directly connected
with the mixed bispectrum associated with these three fields according to the relation \cite{Cooray3} :
\be
C^{XY,Z}_l= \sum_{l_1,l_2} B^{XYZ}_{l l_1 l_2} w_{l_1 l_2} 
\sqrt \frac{(2l_1+1)(2l_2+1)}{4\pi (2l+1)}
\left( \begin{array}{ccc}
l_1 & l_2 & l_3 \\
0 & 0 & 0 \\
\end{array} \right)
\ee
where $w_{l_1 l_2} $ is a filter function that needs to be introduced in a more general approach and represents the spherical transform of the mask.
This power spectrum contains information about all possible triangular configuration when one of the side is fixed at length $l$.

We can now relate the $X$, $Y$ and $Z$ fields introduced above to quantities that we are interested in studying. We therefore expand the observed CMB temperature 
anisotropies $\delta T(\hat{\Omega})$ in terms of the primary anisotropy $\delta T_P$,
due to lensing of primary $\delta T_L$, and the other secondaries generated by the low-redshift large-scale structure $\delta T_S$ :
\be
\delta T(\hat{\Omega})=\delta T_P(\hat{\Omega}) + \delta T_L(\hat{\Omega}) + \delta T_S(\hat{\Omega}) \ .
\ee

Expanding these fields in the Fourier space according to :
\begin{eqnarray}
\delta T_P(\hat{\Omega}) = \sum_{l m} a_{P l m} Y_{l m}(\hat{\Omega}), \nonumber\\
\delta T_L(\hat{\Omega}) = \sum_{l m} \nabla \Theta(\hat{\Omega}) \cdot \nabla T_S(\hat{\Omega}), \\
\delta T_S(\hat{\Omega}) = \sum_{l m} a_{S l m} Y_{l m}(\hat{\Omega}) \nonumber
\end{eqnarray}
we have an expression for the cross-correlation power-spectra which denotes
the coupling of lensing with a specific form of secondary CMB anisotropies. Their bispectrum is given by :
\begin{eqnarray}
B^{PLS}_{l_1 l_2 l_3} = \sum_{m_1 m_2 m_3}
\left( \begin{array}{ccc}
l_1 & l_2 & l_3 \\
m_1 & m_2 & m_3 \\
\end{array} \right) \times \nonumber\\
\times \langle(\delta T_P)_{l_1 m_1}(\delta T_L)_{l_2 m_2}(\delta T_S)_{l_3 m_3} \rangle
\end{eqnarray}
where $(\delta T)_{l m}$ is the anisotropy map expansion to multipole harmonics \cite{Goldberg,Cooray2,SmithZal}.
With explicit calculations, the bispectrum becomes :
\begin{eqnarray}
\label{eq:lensbi}
&&B^{PLS}_{l_1 l_2 l_3} = - \biggl \{ X_{l_3} C_{l_1} \frac{l_2(l_2+1)-l_1(l_1+1)-l_3(l_3+1)}{2} + \nonumber\\
&&+ perm. \biggr \} \sqrt{\frac{(2l_1+1)(2l_2+1)(2l_3+1)}{4 \pi}}
\left( \begin{array}{ccc}
l_1 & l_2 & l_3 \\
0 & 0 & 0 \\
\end{array} \right)
\end{eqnarray}
where $X_{l_3}$ is the lensing potential and secondary anisotropies cross-correlation power spectrum and $C_{l_1}$ is the unlensed power spectrum of CMB anisotropies.

\subsection{Optimised skew spectrum}

Following the discussion in Munshi et al. \cite{Munshi2}, we define a set of 9 different fields  of weighed temperature :
\begin{widetext}
\begin{subequations}
\begin{align}
A_{lm}^1 = \frac{b_l a_{lm}}{\tilde{C}_l b_l^2+Nl} C_l \ \ ; \ \
B_{lm}^1 = \frac{l(l+1)b_l a_{lm}}{\tilde{C}_l b_l^2+Nl} \ \ ; \ \ C_{lm}^1 = X_l \frac{b_l a_{lm}}{\tilde{C}_l b_l^2+Nl} \\
A_{lm}^2 = -\frac{l(l+1)b_l a_{lm}}{\tilde{C}_l b_l^2+Nl} C_l \ \ ; \ \
B_{lm}^2 = \frac{b_l a_{lm}}{\tilde{C}_l b_l^2+Nl} \ \ ; \ \ C_{lm}^2 = X_l \frac{b_l a_{lm}}{\tilde{C}_l b_l^2+Nl} \\
A_{lm}^3 = \frac{b_l a_{lm}}{\tilde{C}_l b_l^2+Nl} C_l  \ \ ; \ \
B_{lm}^3 = \frac{b_l a_{lm}}{\tilde{C}_l b_l^2+Nl} \ \ ; \ \ C_{lm}^3 = -X_l \frac{l(l+1)b_l a_{lm}}{\tilde{C}_l b_l^2+Nl} \ ,
\end{align}
\label{coeff}
\end{subequations}
\end{widetext}
where $b_l$ is the beam transfer function; $N_l$ is the noise power spectrum as obtained from averaging noise maps simulations; $C_l$ and $\tilde{C}_l$ are the unlensed and
the lensed CMB power spectrum, respectively.

From these harmonic coefficients, we also construct 9 sky maps :
\begin{eqnarray}
A^i(\hat{\Omega}) = \sum_{lm} Y_{lm}(\hat{\Omega}) A_{lm}^i, \nonumber\\
B^i(\hat{\Omega}) = \sum_{lm} Y_{lm}(\hat{\Omega}) B_{lm}^i, \\
C^i(\hat{\Omega}) = \sum_{lm} Y_{lm}(\hat{\Omega}) C_{lm}^i \nonumber
\end{eqnarray}
where $i=1,2,3$.

The skewness power spectrum that is weighted to measure non-Gaussianity associated with lensing-secondary correlation can be written as :
%\begin{eqnarray}
\be
C_l^{2,1} =\frac{1}{2l+1} \sum_m \sum_i {\rm Real} \bigg[\bigg( A^i(\hat{\Omega})B^i(\hat{\Omega}) \bigg)_{lm} C^i(\hat{\Omega})_{lm} \bigg] 
%\, . \nonumber \\
\ee
%\end{eqnarray}

The above form is exact for all-sky measurements. To account for partial sky coverage due to the Galactic and foreground mask and inhomogeneous noise,
we also calculate the linear-order correction terms from Ref.~\cite{Munshi2} :
\begin{eqnarray}
C_l^{2,1}&=\frac{1}{f_{sky}} \sum_i \bigg[ C_l^{AB,C} - C_l^{A<B,C>} - C_l^{B,<A,C>} + \nonumber\\
& - C_l^{<AB>,C} \bigg]^i  \, .
\label{est_skew}
\end{eqnarray}
The term above without an averaging is the direct estimate from data while the averaged corrective terms such as
$C_l^{A<B,C>}$ are obtained by cross-correlating the product of the observed A map with the simulated B and C maps and then taking
an ensemble average over many realizations. The reduction in the sky are due to mask 
is corrected dividing by the observed sky fraction $f_{sky}$.

As discussed in Ref.~\cite{Munshi2}, it is possible to show that this quantity is directly related to the bispectrum :
\be
C_l^{2,1} = \frac{1}{2 l +1} \sum_{l l_1 l_2} \frac{\hat{B}_{l l_1 l_2} \left(B^{PLS}_{l l_1 l_2 }\right)^c}{\tilde C_l \tilde C_{l_1} \tilde C_{l_2}} \, ,
\ee
where $\hat{B}_{l l_1 l_2}$ is the total bispectrum in the data and $\left(B^{PLS}_{l l_1 l_2 }\right)^c$ is the shape of the bispectrum that
we have employed by weighting the A, B and C maps. This bispectrum is equal to the form written in equation~(\ref{eq:lensbi}), with permutations only restricted to $l_1 \rightarrow l_2$
while $l_3$ is kept fixed to $X_{l_3}$. 

We assume the total bispectrum present in the data is a combination of the both the
lensing-secondary bispectrum and contaminations such as point sources. When fitting to measurements, we will construct the map $C^i$ in above
by appropriately weighting it with $X_l$ to study the cross-correlation of lensing potential with SZ and ISW separately.
We assume  that the bispectrum in the data is composed by these two effects with two unknown amplitudes relative to the prediction
under the fiducial cosmological model. The comparison between the data and the modeled expectation will be used to determine
the two relative

\section{Data analysis}
We summarize our analysis in the following steps:

\subsection{Data and Simulations}

We have considered the WMAP 5-year Stokes-I raw and clean sky maps for the Q, W and V frequency bands, obtained from the public lambda website\footnote{http://lambda.gsfc/nasa.gov}.
We use the Healpix's synfast code \cite{healpix} to generate $250$ CMB temperature anisotropy Gaussian maps giving in input the WMAP 5-year best-fit CMB anisotropy 
power spectrum with cosmological parameters: $H_0 = 71.9 \  \rm{km/s/Mpc}$, $\Omega_b h^2 = 0.02273$,
$\Omega_c h^2 = 0.1099$, $n_s = 0.963$ and $\tau = 0.087$. We require $Nside=512$ and a maximum multipole equal to $600$.

In the same way, we generate $250$ noise maps with noise properties consistent with WMAP Q, W and V frequency bands :
\be
N(\hat{\Omega}) = \frac{\sigma_0}{\sqrt{N_{obs}}} n(\hat{\Omega})
\ee
where $N(\hat{\Omega})$ is the final noise map obtained from a white noise map $n(\hat{\Omega})$  combined with the WMAP rms
noise per observation, $\sigma_0$, and the number of observations per pixel, $N_{obs}$. We extract $N_{obs}$ from 
the WMAP 5-year Stokes-I sky maps fits files and take $\sigma_0 = 2.197, 3.133, 6.538 \ \rm{mK}$ for the Q, V and W 5-year data, respectively, as published on the lambda website by the WMAP team.

We analyze both raw maps as well as foreground-cleaned maps provided by the WMAP team. We show and tabulate results separately for these two options.

We use the Healpix anafast code \cite{healpix} and the $KQ75$ mask to extract the multipole coefficients for each frequency band out to $l_{max}=600$ for WMAP maps, $a_{lm}^D$, 
simulated Gaussian maps, $a_{lm}^G$, and simulated noise maps, $a_{lm}^N$.
The noise spectrum needed for computing the denominators in (\ref{coeff}) is obtained averaging the simulated noise spectra over the solid angle and considering the sky-cut according to 
the relation :
\be
N_l = \Omega \int \frac{d^2 \hat{\boldsymbol{n}} \ \sigma_0^2 \ M(\hat{\boldsymbol{n}})}{4 \pi f_{sky} N_{obs}(\hat{\boldsymbol{n}})} \, ,
\ee
where $\Omega \equiv 4 \pi / N_{pixel}$ is the solid angle per pixel, $M(\hat{\boldsymbol{n}})$ is the $KQ75$ mask and
$f_{sky}=0.718$ is the corresponding observed sky fraction. 

We calculate the lensed and unlensed CMB power spectrum with the public CAMB code \cite{camb}
using again the cosmological parameters from the WMAP 5-year best fit model.

We put everything together to obtain all coefficients in (\ref{coeff}) and the relative sky maps considering $a_{lm}= a_{lm}^D$ for data instead, in the case of simulations, we 
need to consider noise and beam contribution to multipoles: $a_{lm} = a_{lm}^G b_l + a_{lm}^N$, so our gaussian multipoles are convolved with the frequency-dependent beam 
transfer function $b_l$ and added to the noise multipoles.

\subsection{Skewness power spectrum}

\begin{figure}[htb!]
\centerline{\includegraphics[width=9.1cm]{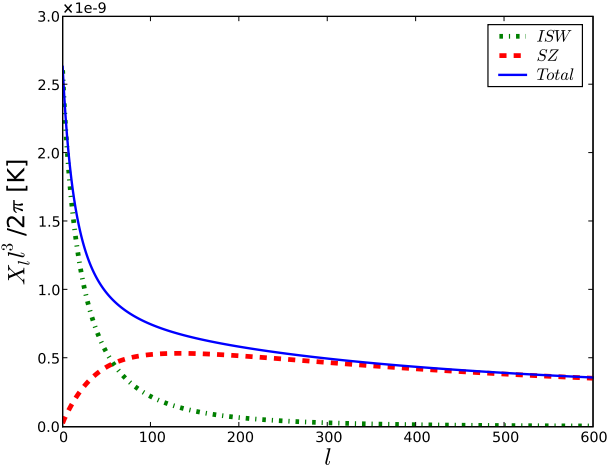}}
\caption{\label{xl} The lensing-secondary anisotropy cross-correlation power spectrum. We consider two secondary effects here with ISW and SZ as these are the two dominant lensing-secondary correlations expected. The blue solid line is the total contribution when we consider both ISW and SZ effects.}
\end{figure}

We estimate the $C_l^{2,1}$ according to equation (\ref{est_skew}) for each frequency band and for different
lensing-secondary anisotropy cross-correlation power spectrum; in particular, $X_l$ ISW is the spectrum of cross-correlating lensing with Integrate Sachs-Wolfe effect 
\cite{Sachs} and, in the same way, $X_l$ SZ  for Sunyaev-Zel'dovich effect \cite{Sunyaev} (see Figure \ref{xl}).
We calculate these in the fiducial cosmological model consistent with WMAP 5-year data and making use of the halo model approach to describe the SZ signal  \cite{CooShe,Cooray4,KS}.
The ISW effect is described through the standard linear power spectrum of the potential field and the CMB lensing potential is also modeled using the linear fluctuations 
\cite{Afshordi,Cooray5}.

\begin{figure}[htb!]
\centerline{\includegraphics[width=9.4cm]{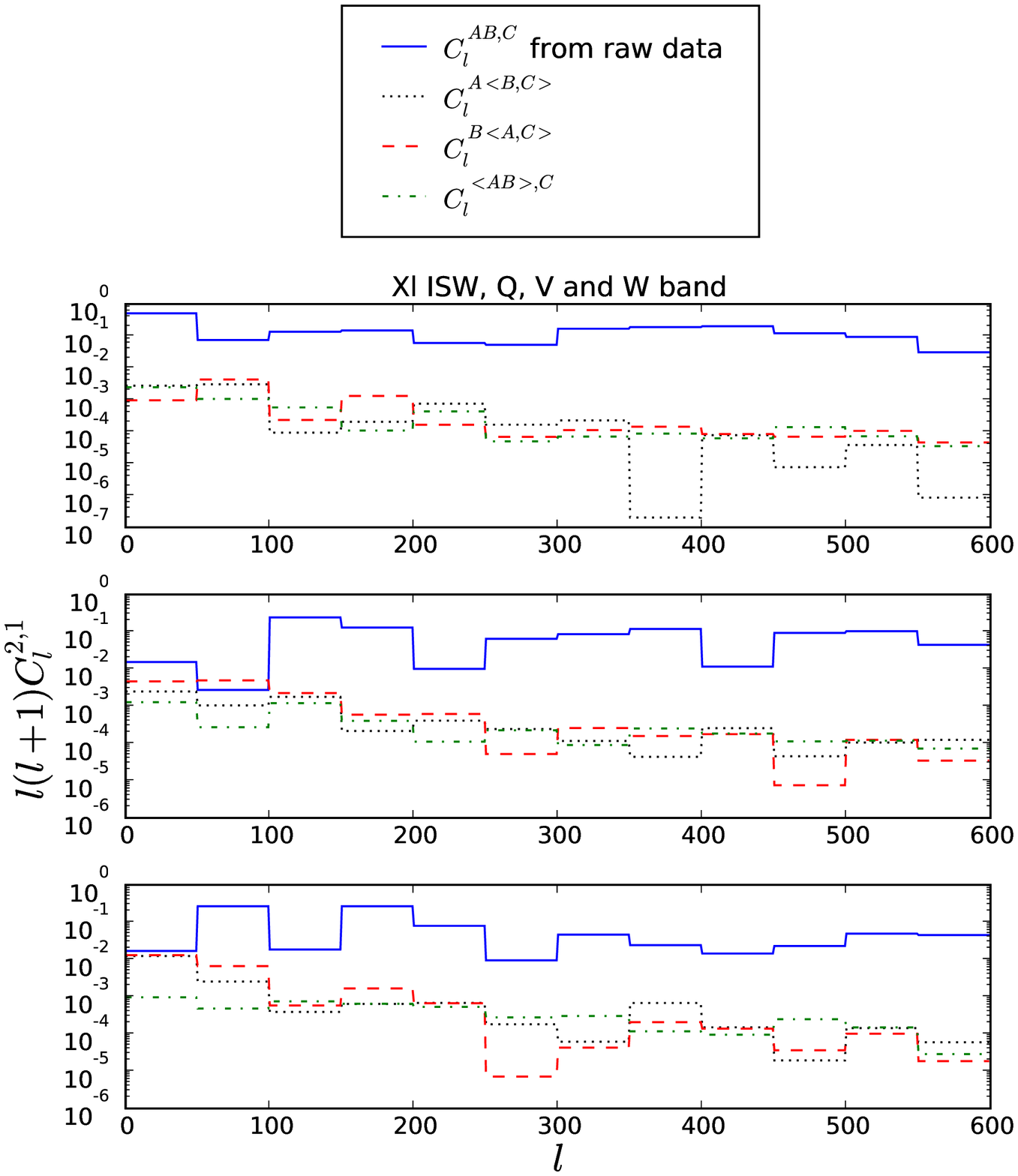}}
\centerline{\includegraphics[width=9.4cm]{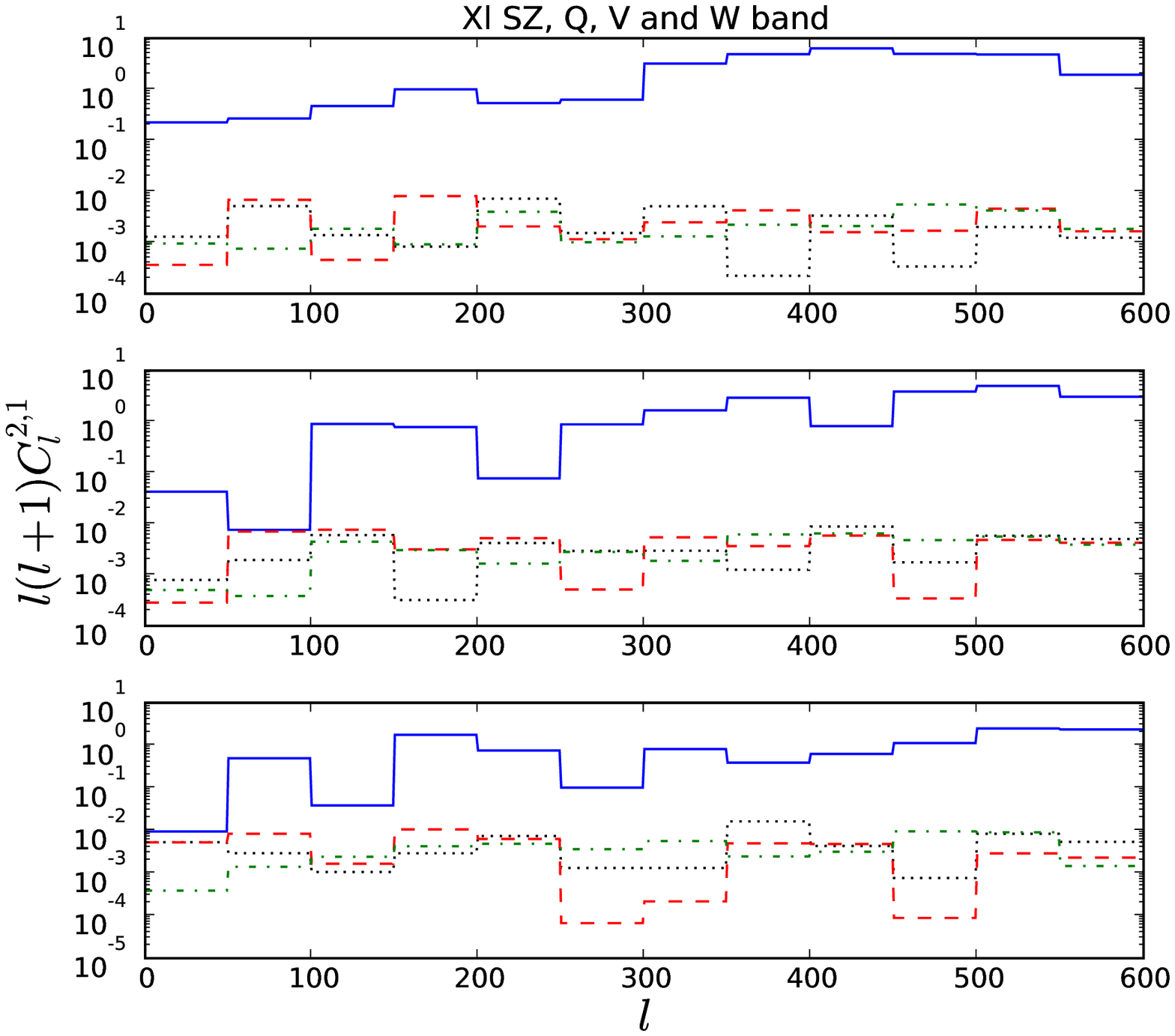}}
\caption{\label{corrections} Corrective terms compared with the raw data estimator for Q, V, and W band 
for $X_l$ with ISW (top) and SZ (bottom), respectively.}
\end{figure}

\begin{figure}[htb!]
\centerline{\includegraphics[width=9.4cm]{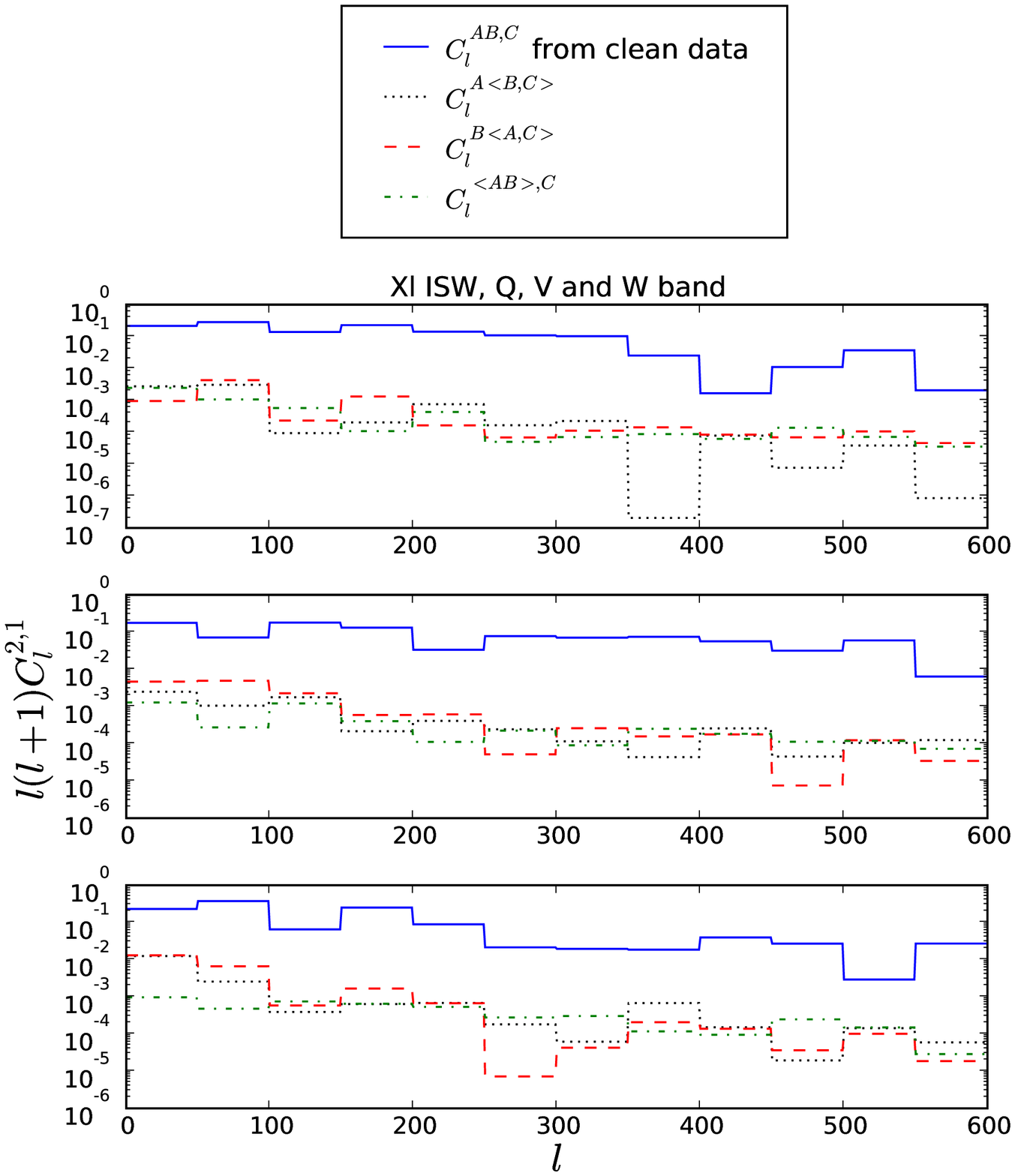}}
\centerline{\includegraphics[width=9.4cm]{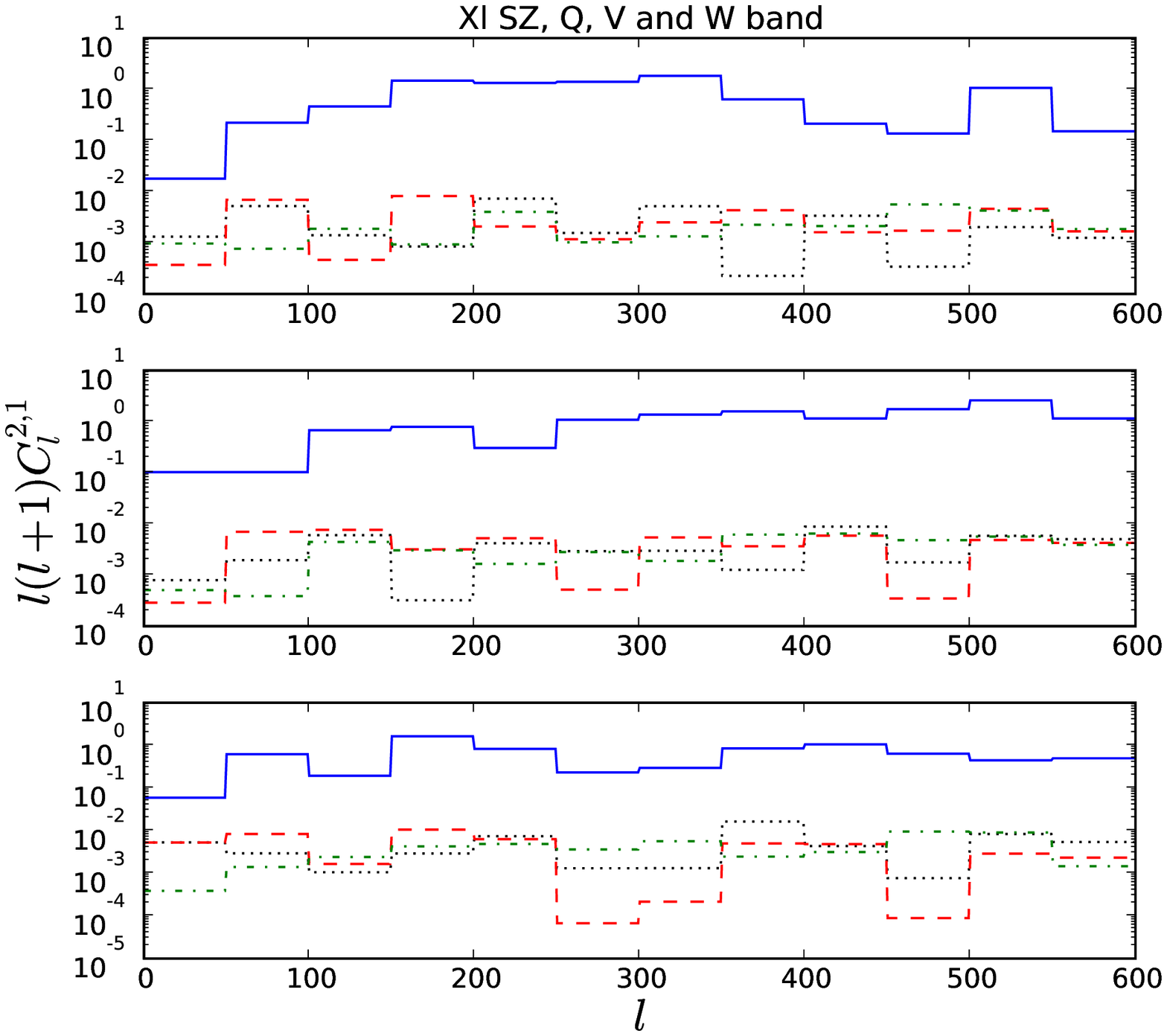}}
\caption{\label{corrections_clean} Corrective terms compared with the clean data estimator for Q, V, and W band 
for $X_l$ with ISW (top) and with SZ (bottom), respectively.}
\end{figure}

In Figures \ref{corrections} and \ref{corrections_clean} we show all terms of equation (\ref{est_skew}). It's evident that linear terms are not significant compared to the others coming from data only. However, we are using all the contributions when calculating the skewness spectrum.

\begin{figure}[!htb]
\centerline{\includegraphics[width=8.6cm]{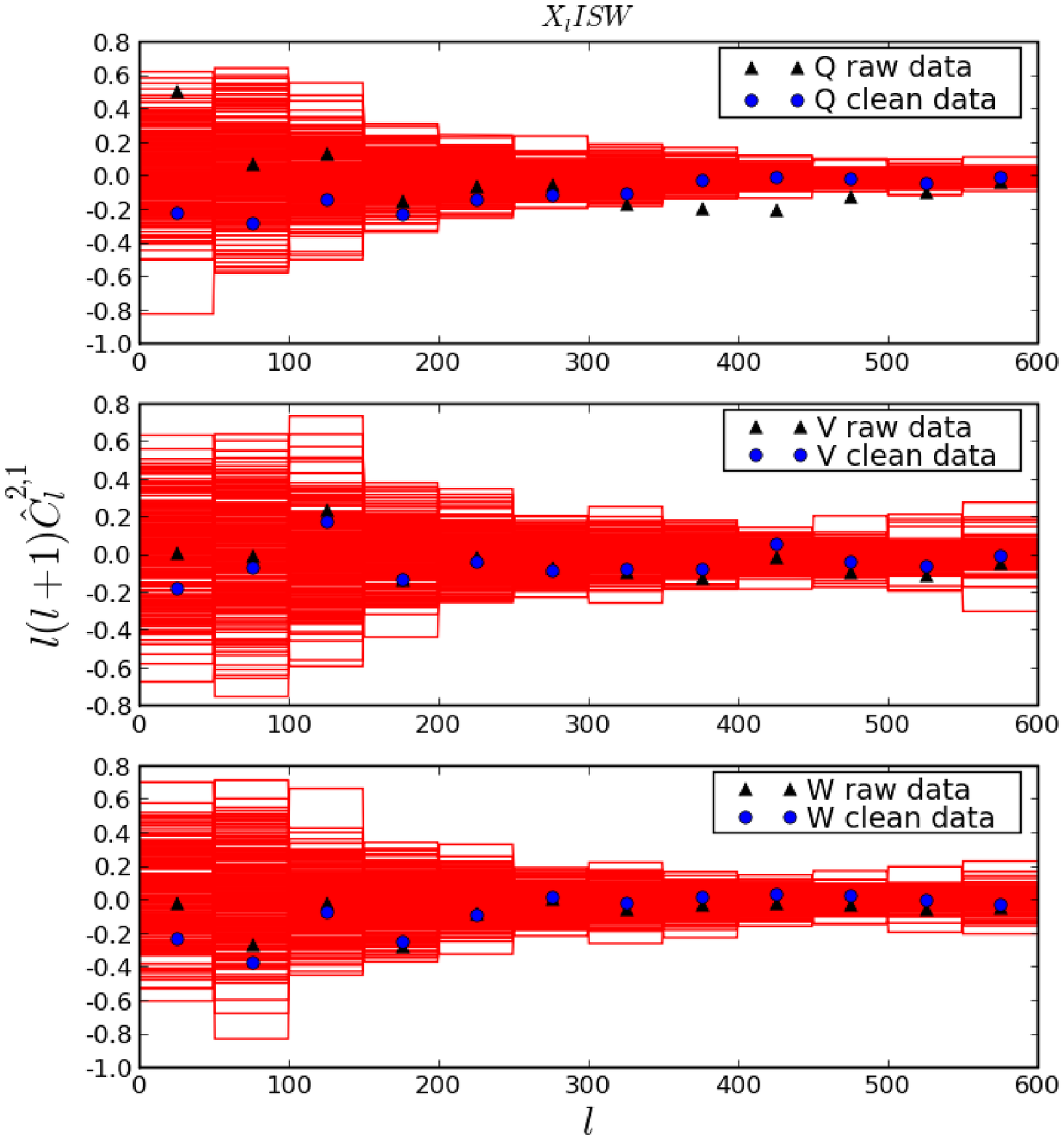}}
\centerline{\includegraphics[width=8.6cm]{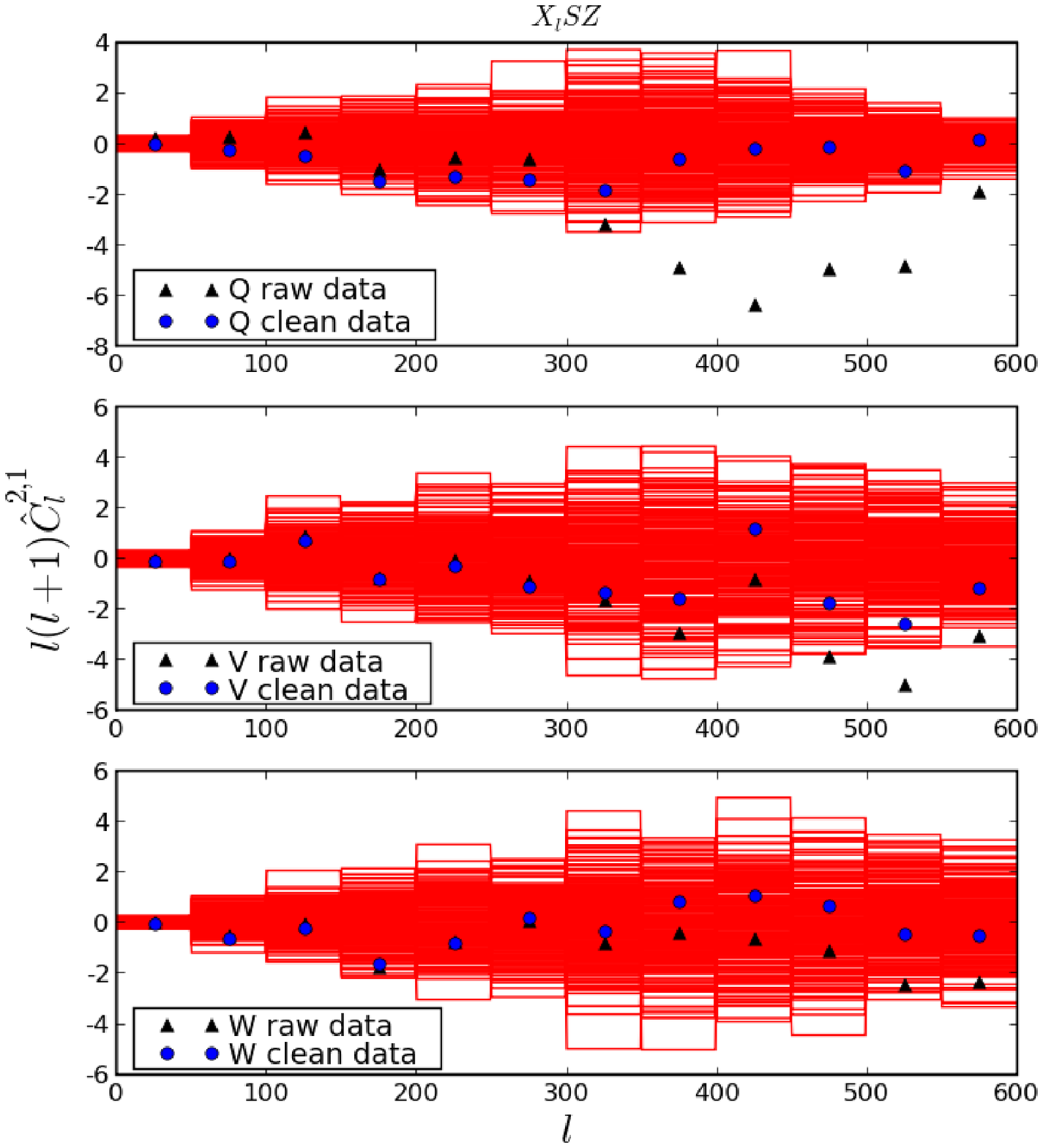}}
\caption{\label{binned} Binned data obtained calculating all the corrective terms compared to the skew spectrum of simulations for $X_l$ ISW (top) and $X_l$ SZ (bottom).}
\end{figure}

In Figure \ref{binned} we report both raw and foregrounds-cleaned maps data with all corrective terms considered compared to $C_l^{2,1}$ from Gaussian simulations.

\subsection{Best Fit Parameter Estimation}

\subsubsection{Secondary non-Gaussianity only}

At this stage, we make an assumption, relaxed later, that there is no primordial non-Gaussianity.

\begin{figure}[htb!]
\centerline{\includegraphics[width=10cm]{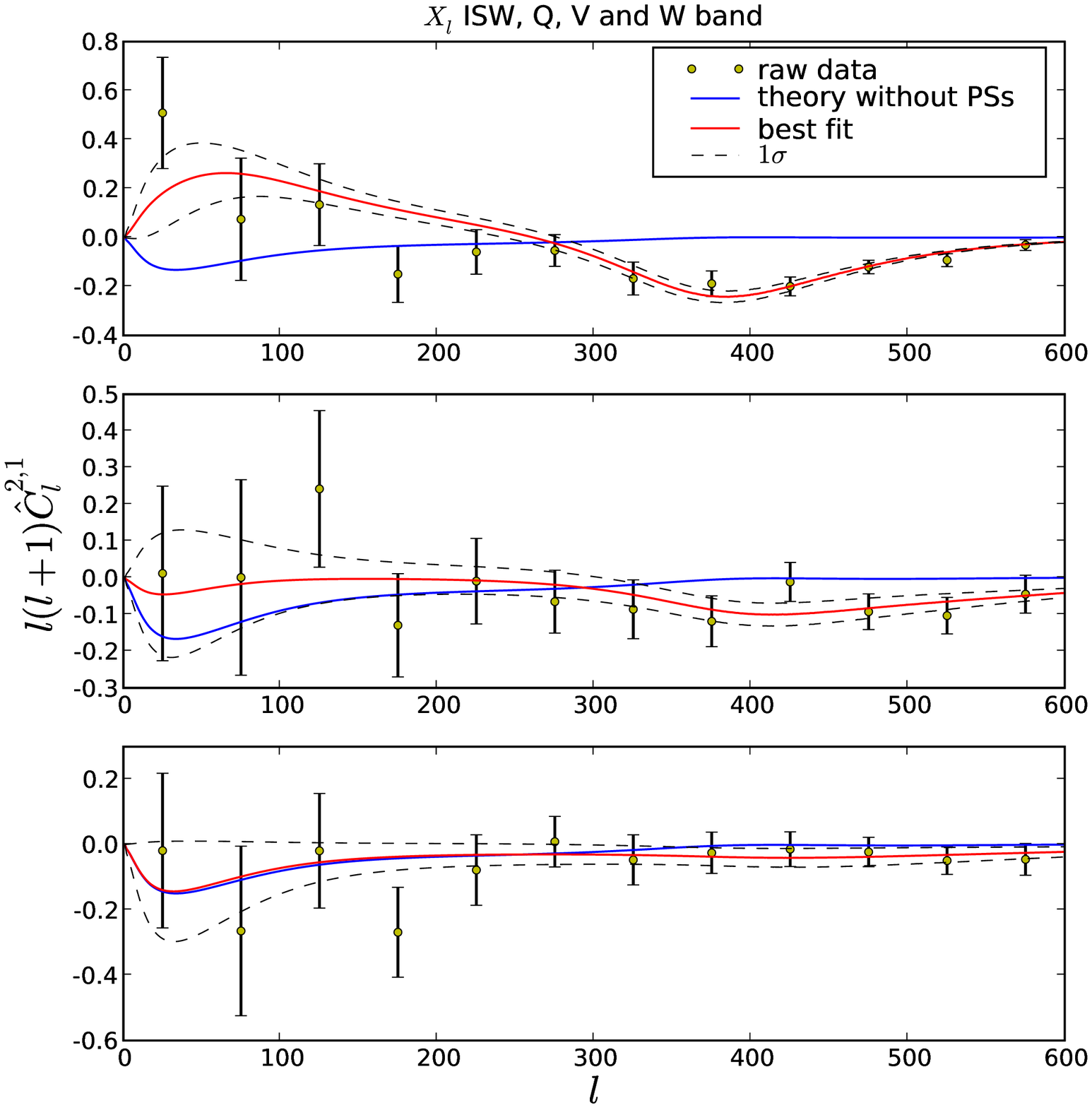}}
\centerline{\includegraphics[width=10cm]{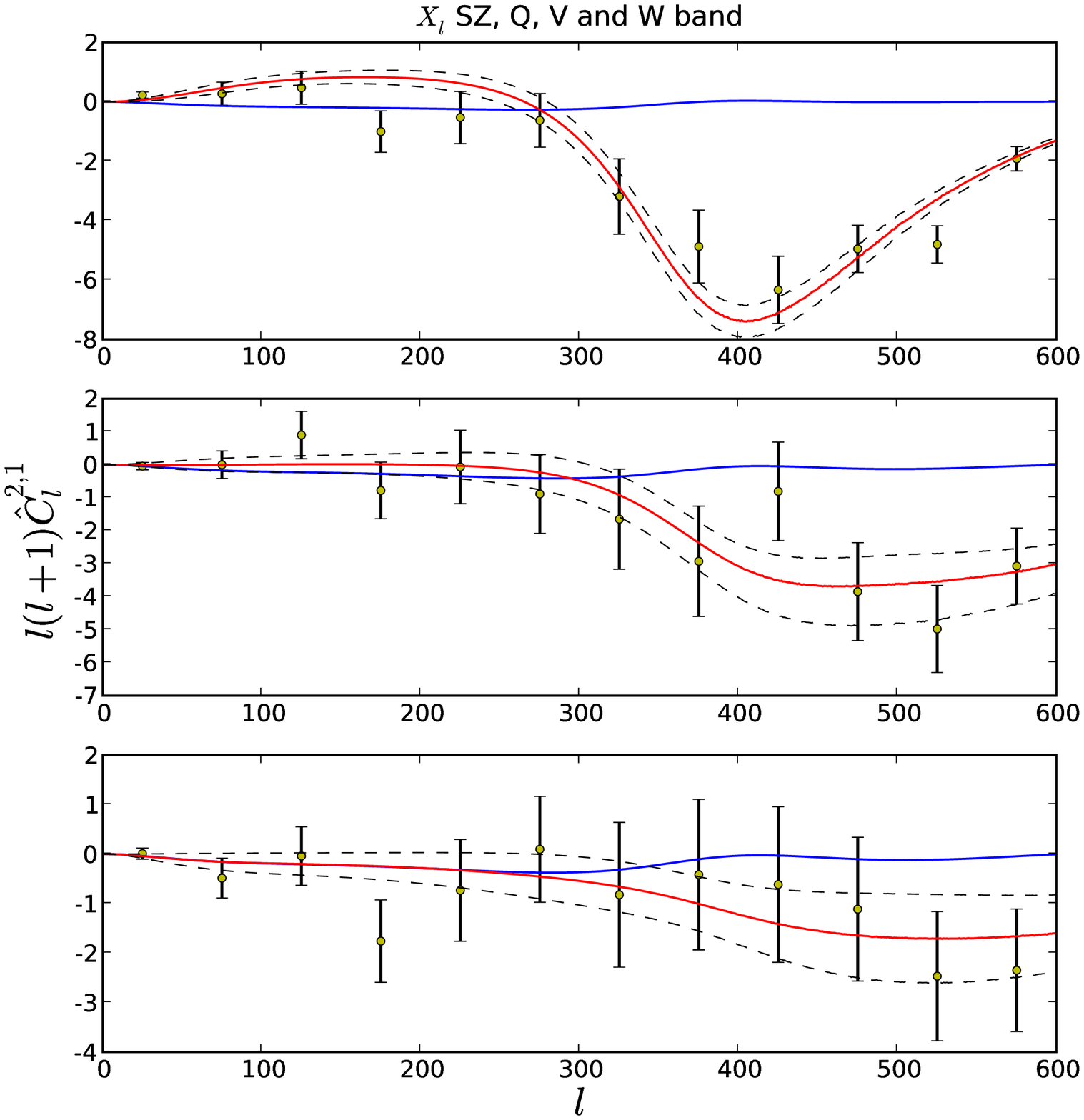}}
\caption{\label{fit} Q, V and W band WMAP raw data with error bars from simulations variance compared with the
 theoretical models and the best fit results for $X_l$ ISW (top)
and $X_l$ SZ (bottom).}
\end{figure}

\begin{figure}[htb!]
\centerline{\includegraphics[width=10cm]{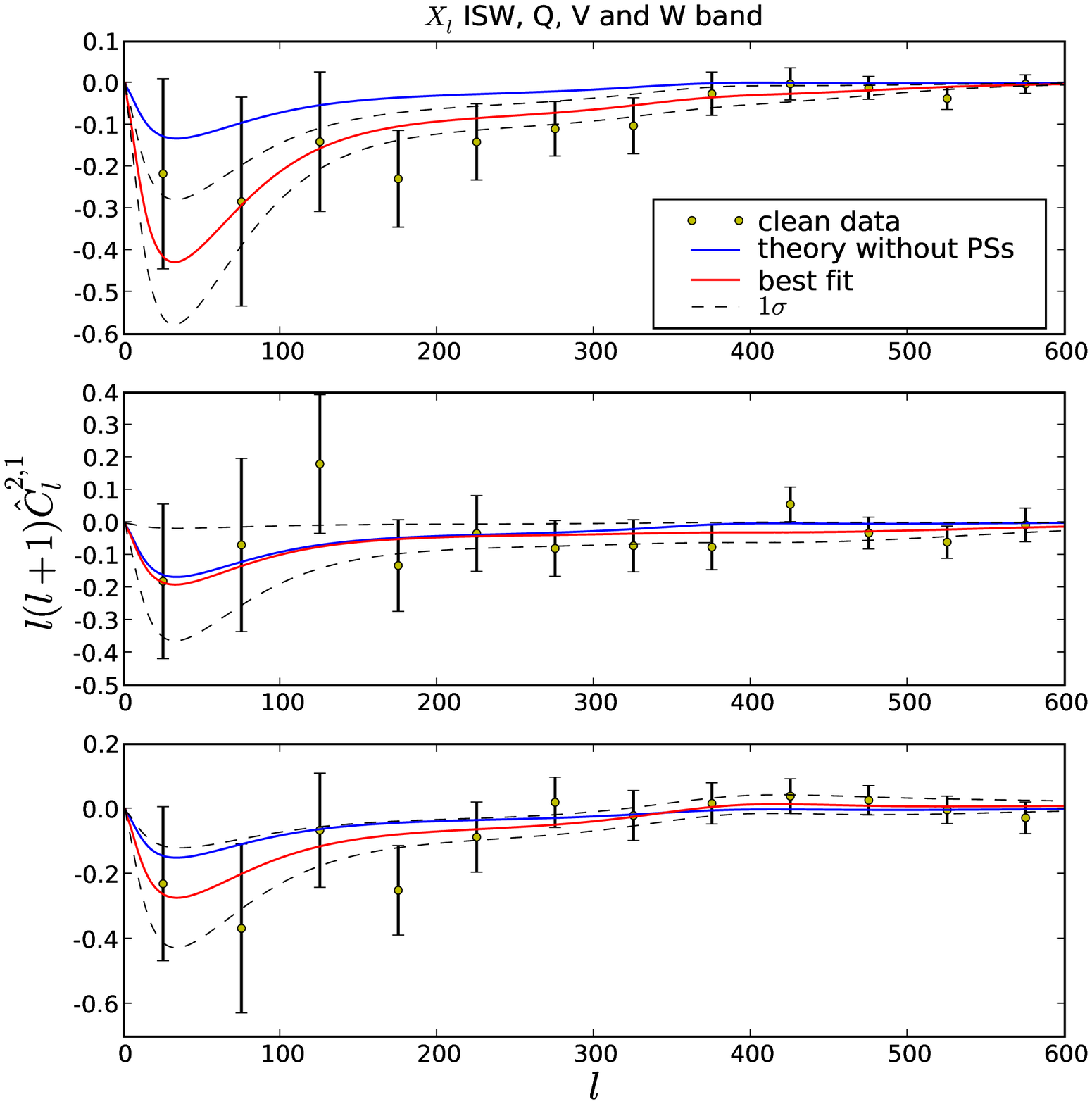}}
\centerline{\includegraphics[width=10cm]{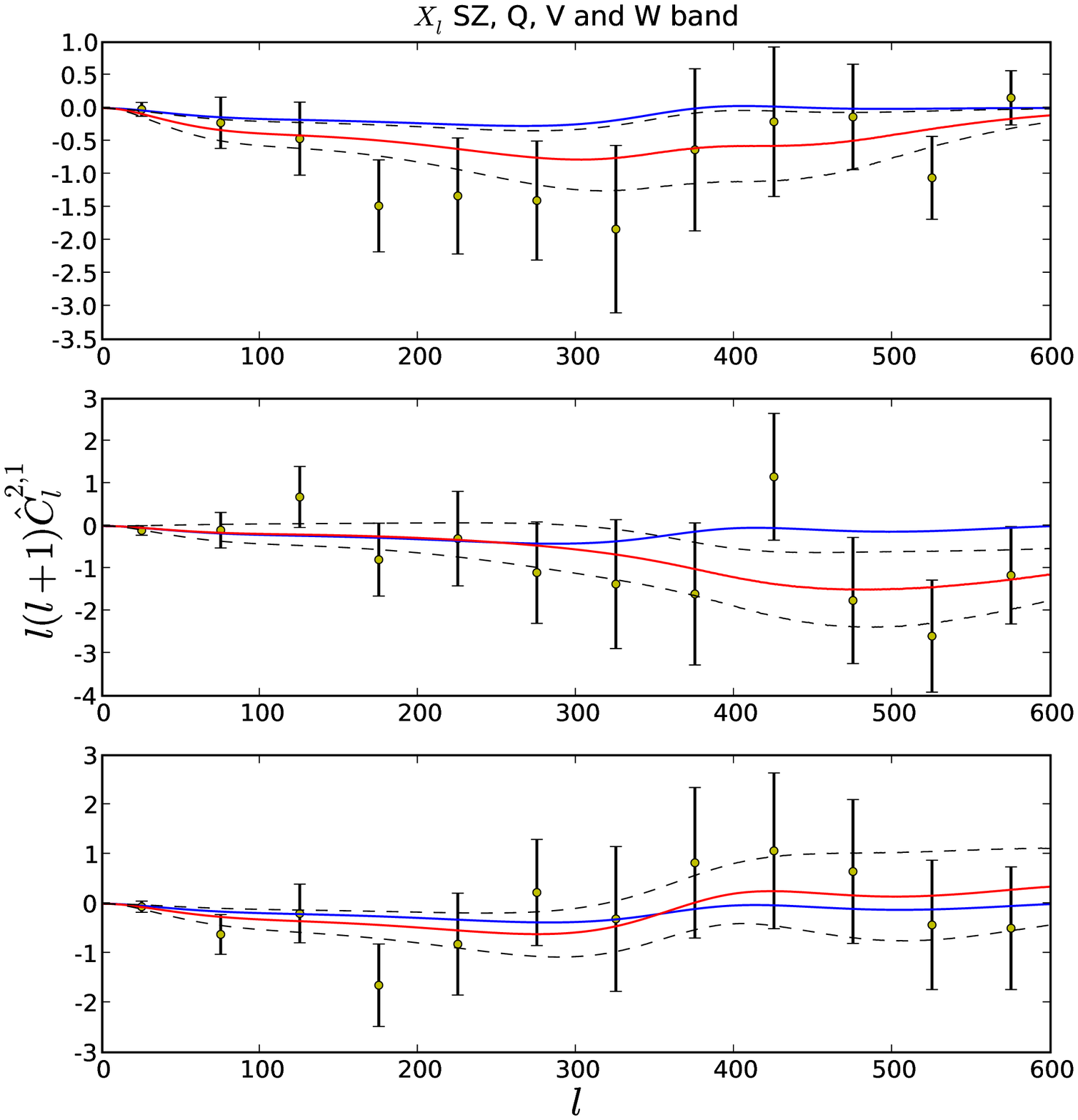}}
\caption{\label{fit_clean} Q, V and W band WMAP clean data with error bars from simulations variance compared with the
 theoretical models and the best fit results for $X_l$ ISW (top)
and $X_l$ SZ (bottom).}
\end{figure}

In Figures \ref{fit} and \ref{fit_clean} we plot binned data for each $X_l$ configuration
and for each frequency band with a $\Delta l = 50$. The corresponding error bars are from simulations 
variance and the solid blue line represents a calculation from the halo model for the lensing-SZ and lensing-ISW
correlations. These calculations are described in Ref.~\cite{Munshi2}.

In making these estimates, we also allow for unresolved point sources. We calculated the overlap between the lensing-SZ
and lensing-ISW estimators with the point sources given by their shot-noise term of the angular bispectrum. We parameterize
point source amplitude such that :
\be
B^{\rm PS}_{l_1l_2l_3} = b_{\rm src} \sqrt \frac{(2l_1+1)(2l_2+1)}{4\pi (2l+1)}
\left( \begin{array}{ccc}
l_1 & l_2 & l_3 \\
0 & 0 & 0 \\
\end{array} \right) \ ,
\label{ps}
\ee
and we consider $b_{\rm src}= B_i \times 10^{-25}$ sr$^{2}$ as a function of frequency $i$.

In order to estimate the relative amplitudes of SZ and ISW correlation with the lensing potential as well as the contaminant 
contribution from unresolved point sources, we compare the data to a model calculation that contains lensing bispectra and the overlap between lensing and point source bispectra :
\be
{\bf D} = A \times C_l^{2,1sec-lens} + B_i \times E_l^{2,1 PS} \ ,
\label{datafit}
\ee
where :
\be
E_l^{2,1} = \frac{1}{2 l +1} \sum_{l l_1 l_2} \frac{\hat{B}^{\rm PS}_{l l_1 l_2} \left(B^{PLS}_{l l_1 l_2 }\right)^c}{\tilde C_l \tilde C_{l_1} \tilde C_{l_2}} \,,
\ee
and $\hat{B}^{\rm PS}_{l l_1 l_2}$ is the point source shot-noise bispectrum described in equation~(\ref{ps}).
The parameter $A$ denotes the total amplitude of the combined lensing-SZ and lensing-ISW correlations.

In doing the above model fit to multi-frequency measurements of the 2,1 correlator optimized for lensing-secondary correlations,  
we assume that the lensing-secondary cross-correlation amplitude is
 frequency-independent, except for the known frequency dependence of the SZ effect
taken as part of the model calculation of $X_l$. We assume three different amplitudes for point sources in each of Q, V, and W bands of WMAP.

Our technique as implemented currently does not allow us to separate the lensing-SZ correlation amplitude from
lensing-ISW amplitude as the two are found to be highly degenerate with each other. In future, especially with Planck,
one may be able to separate lensing-SZ from lensing-ISW based on the SZ frequency dependence.

To estimate the amplitudes $A$ and $B_i$ we define a $\chi^2$ merit function :
\be
\chi^2= (\bs{y}^T- \bs{T}\cdot \bs{P})^T \bs{C}^{-1}(\bs{y}-\bs{T} \cdot \bs{P})
\ee
where $\bs{y}$ is the data vector (binned $C_l^{2,1}$ for each frequency band and $X_l$ configuration),
 $\bs{T}$ is the theory matrix and $\bs{P}=(A,B_i)$ is the parameter vector that we want to estimate. $\bs{C}$ is the covariance matrix from simulations.

To determine the parameters, we minimize the $\chi^2$ function explicitly and obtain the best-fit parameters as :
\be
\bs{P}=(\bs{T}^T \bs{C}^{-1} \bs{T})^{-1} (\bs{T}^T \bs{C}^{-1} \bs{y})
\ee
and the error bars are obtained by the diagonal elements of the following matrix :
\be
\sigma_{\bs{P}}^2 = (\bs{T}^T \bs{C}^{-1} \bs{T})^{-1} \ .
\ee

We summarize results related to the amplitude determination for different frequency bands 
in the tables \ref{par} and \ref{par_clean} for raw and foreground-cleaned maps respectively.
We categorize our results by the choice we make in setting $X_l$ while generating $C$ maps. Here, $X_l$ forms a template to
search for the lensing-secondary cross-correlation. In $C_l^{2,1}$, due to weighting, one of the bispectra probes the total non-Gaussianity in CMB data that could come from a combination of effects from primordial non-Gaussianity to
lensing and radio point sources. The second bispectrum forces a certain configuration and the overlap between the total
and the assumed shape of the bispectrum determines the shape of $C_l^{2,1}$ measured from the data.
By setting a function for $X_l$, we set the overall normalization of the prescribed lensing-secondary cross-correlation bispectrum.
Thus the amplitude $A$ we determine from the data is simply the overall amplitude of the non-Gaussianity associated with the
overlap between lensing-secondary cross-correlation and all forms of non-Gaussianities that are present in the data.
We remove the confusion associated with point sources, which is expected to be significant, by explicitly calculating the
overlap between lensing-(SZ+ISW) bispectrum and the shot-noise form of the point source bispectrum.

After accounting for the confusion from point sources generated by the overlap of the point source shot-noise bispectrum and the
lensing-secondary anisotropy cross-correlation bispectrum, we find no significant detection of the lensing effect in existing WMAP data.
We constrain the overall normalization of the lensing-SZ and lensing-ISW angular cross-power spectra to be 
0.42 $\pm$ 0.86 and 1.19 $\pm$ 0.86 in combined V and W-band raw and foreground-cleaned maps provided by the WMAP team, respectively.
The point source amplitude we determine from the raw map of Q-band with $b_{\rm src}=(67.8 \pm 5.4)\times 10^{-25}$ sr$^2$  is higher than
the estimate by the WMAP team with $(6.0 \pm 1.3) \times 10^{-5}$ $\mu$K$^3$-sr$^2$ \cite{Komatsu} (the value we determine is $(13.7 \pm 1.1) \times 10^{-5}$
$\mu$K$^3$-sr$^2$ in the same units used by the WMAP team).  We find similarly a factor of 2 increase in V-band map as well.

In the case of clean maps, we find $b_{\rm src}=(6.2 \pm 5.4)\times 10^{-25}$ sr$^2$, which is smaller than the
WMAP team's estimate with clean maps for the Q band with  
$(4.3 \pm 1.3) \times 10^{-5}$ $\mu$K$^3$-sr$^2$ \cite{Komatsu} (the value we determine is $(1.4 \pm 1.1) \times 10^{-5}$
$\mu$K$^3$-sr$^2$ in the same units used by the WMAP team). We find similar differences in V and W-band as well.

It is unclear exactly where these differences come from. We do not employ the same E-statistic that is optimized for point
sources as the WMAP team in the present study. 

\subsubsection{Inclusion of primordial non-Gaussianity}

To study the impact of primordial non-Gaussianity we now fit the data by modifying equation (\ref{datafit}) to include
a local form of non-Gaussianity with amplitude $f_{\rm NL}$ :
\be
{\bf D} = A \times C_l^{2,1sec-lens} + B_i \times E_l^{2,1 PS} +f_{\rm NL} Y_l^{2,1 prim}\ ,
\ee
where now :
\be
Y_l^{2,1} = \frac{1}{2 l +1} \sum_{l l_1 l_2} \frac{\hat{B}^{prim}_{l l_1 l_2} \left(B^{PLS}_{l l_1 l_2 }\right)^c}{\tilde C_l \tilde C_{l_1} \tilde C_{l_2}} \,,
\ee
involves the overlap between lensing-secondary and primordial non-Gaussianities with the overall amplitude of the primordial non-Gaussianity determined by
$f_{\rm NL}$ \cite{Smidt,Smith}.  

Including primordial non-Gaussianity confuses 
the detection of lensing-secondary correlations and leads to a factor of 2 degradation in the error of the amplitude of lensing-secondary correlation
power spectrum (see Table~III).  The estimator as constituted is not optimised to detect primordial non-Gaussianity, and we find a rather weaker limit of $f_{\rm NL}=-13\pm 62$ from the clean V+W maps, with a larger error bar than in the study of Ref.~\cite{Smidt} which uses
the optimised estimator of Ref.~\cite{Munshi1} for primordial non-Gaussianity specifically.  

We emphasize that our current
study is more focused towards a detection of the lensing-secondary correlation in WMAP data.
In an upcoming paper, we will present a combined analysis of three estimators of the 2,1 correlator optimized separately
for primordial non-Gaussianity, point sources, and lensing effects.

\begin{table*}[!htb]
\begin{center}
\begin{tabular}{@{\ \ }c @{\ \ }|@{\ \ }c@{\ \ }|@{\ \ }c@{\ \ }|@{\ \ }c@{\ \ }|@{\ \ }c@{\ \ }|@{\ \ }c@{\ \ }|@{\ \ }c @{\ \ }}
Frequency & $X_l$ & $A$ & $B_Q$ & $B_V$ & $B_W$ & $\chi^2/dof$ \\
\hline
Q & SZ+ISW & $-1.59 \pm 1.21$ & $67.8 \pm 5.4$ & & & $1.67$\\
V & & $0.06 \pm 1.08$ & &  $11.4 \pm 2.4$ & & $0.85$ \\
W & & $1.01 \pm 1.06$  & & & $5.4 \pm 2.6$ & $0.58$\\
Q+W+V & & $0.07 \pm 0.82$ & $67.8 \pm 5.2$ &$12.4 \pm 2.2$& $5.8 \pm 2.4$&  $1.22$ \\
W+V & & $0.42 \pm 0.86$ & & $11.8 \pm 2.2$ & $5.2 \pm 2.6 $ & $0.82$\\
\hline
Q & ISW & $-1.03 \pm 1.19$ & $123.4 \pm 12.4$ & & & $1.02$\\
V & & $0.33 \pm 1.06$ & & $20.8 \pm 6.2$ & &$0.67$\\
W & & $0.99 \pm 1.05$ & & &$10.0 \pm 7.0$ & $0.50$\\
Q+W+V & & $0.43 \pm 0.82$& $128.2 \pm 11.8$ & $22.6 \pm 5.8$ & $11.6 \pm 6.6$ & $1.02$  \\
W+V & & $0.48 \pm 0.86$ & & $23.2 \pm 5.8$ & $8.4 \pm 6.8$ & $0.66$\\
\hline
Q & SZ & $-1.47 \pm 1.33$ & $136.0 \pm 10.8$ & & & $1.68$\\
V & & $0.24 \pm 1.18$ & & $22.8 \pm 4.6$ & & $0.84$\\
W & & $1.09 \pm 1.14$ & & &$ 10.8 \pm 5.2$ & $0.60$\\
Q+W+V & &$0.13 \pm 0.89$ & $136.2 \pm 10.4$ & $25.2 \pm 4.2$ & $12.2 \pm 5.0$ & $1.23$  \\
W+V & & $0.57 \pm 0.94$ & & $23.8 \pm 4.4$ & $10.4 \pm 5.0$ & $0.85$\\
\end{tabular}
\caption{Amplitude parameters estimation using WMAP raw maps.}
\label{par}
\end{center}
\end{table*}

\begin{table*}[!htb]
\begin{center}
\begin{tabular}{@{\ \ }c @{\ \ }|@{\ \ }c@{\ \ }|@{\ \ }c@{\ \ }|@{\ \ }c@{\ \ }|@{\ \ }c@{\ \ }|@{\ \ }c@{\ \ }|@{\ \ }c @{\ \ }}
Frequency & $X_l$ & $A$ & $B_Q$ & $B_V$ & $B_W$ & $\chi^2/dof$ \\
\hline
Q & SZ+ISW & $2.93 \pm 1.21$ & $ 6.2\pm 5.4$ & & & $0.55$\\
V & & $0.93 \pm 1.08$ & &  $4.4 \pm 2.4$ & & $0.70$ \\
W & & $1.73 \pm 1.06$  & & & $-1.3 \pm 2.6$ & $0.46$\\
Q+W+V & & $1.56 \pm 0.82$ & $8.8 \pm 5.2$ &$4.2 \pm 2.2$& $-1.4 \pm 2.4$&  $0.82$ \\
W+V & & $1.19 \pm 0.86$ & & $5.0 \pm 2.2$ & $-1.6 \pm 2.6 $ & $0.77$\\
\hline
Q & ISW & $3.32 \pm 1.19$ & $17.2 \pm 12.6$ & & & $0.34$\\
V & & $1.16 \pm 1.06$ & & $5.8 \pm 6.4$ & &$0.62$\\
W & & $1.81 \pm 1.05$ & & &$-4.3 \pm 7.0$ & $0.42$\\
Q+W+V & & $1.76 \pm 0.82$& $25.4 \pm 11.8$ & $5.6 \pm 5.8$ & $-5.4 \pm 6.6$ & $0.86$  \\
W+V & & $1.33 \pm 0.86$ & & $7.8 \pm 6.0$ & $-6.0 \pm 6.8$ & $0.67$\\
\hline
Q & SZ & $2.58 \pm 1.33$ & $12.2 \pm 10.8$ & & & $0.62$\\
V & & $0.99 \pm 1.18$ & & $8.7 \pm 4.7$ & & $0.70$\\
W & & $1.66 \pm 1.14$ & & &$ -2.5 \pm 5.2$ & $0.49$\\
Q+W+V & &$1.47 \pm 0.89$ & $16.6 \pm 10.4$ & $8.2 \pm 4.4$ & $-2.2 \pm 5.0$ & $0.69$  \\
W+V & & $1.22 \pm 0.93$ & & $10.0 \pm 4.4$ & $-3.0 \pm 5.0$ & $0.80$\\
\end{tabular}
\caption{Amplitude parameters estimation using WMAP foreground cleaned maps.}
\label{par_clean}
\end{center}
\end{table*}

\begin{figure*}[htb!]
\centering
\begin{tabular}{cc}
\epsfig{file=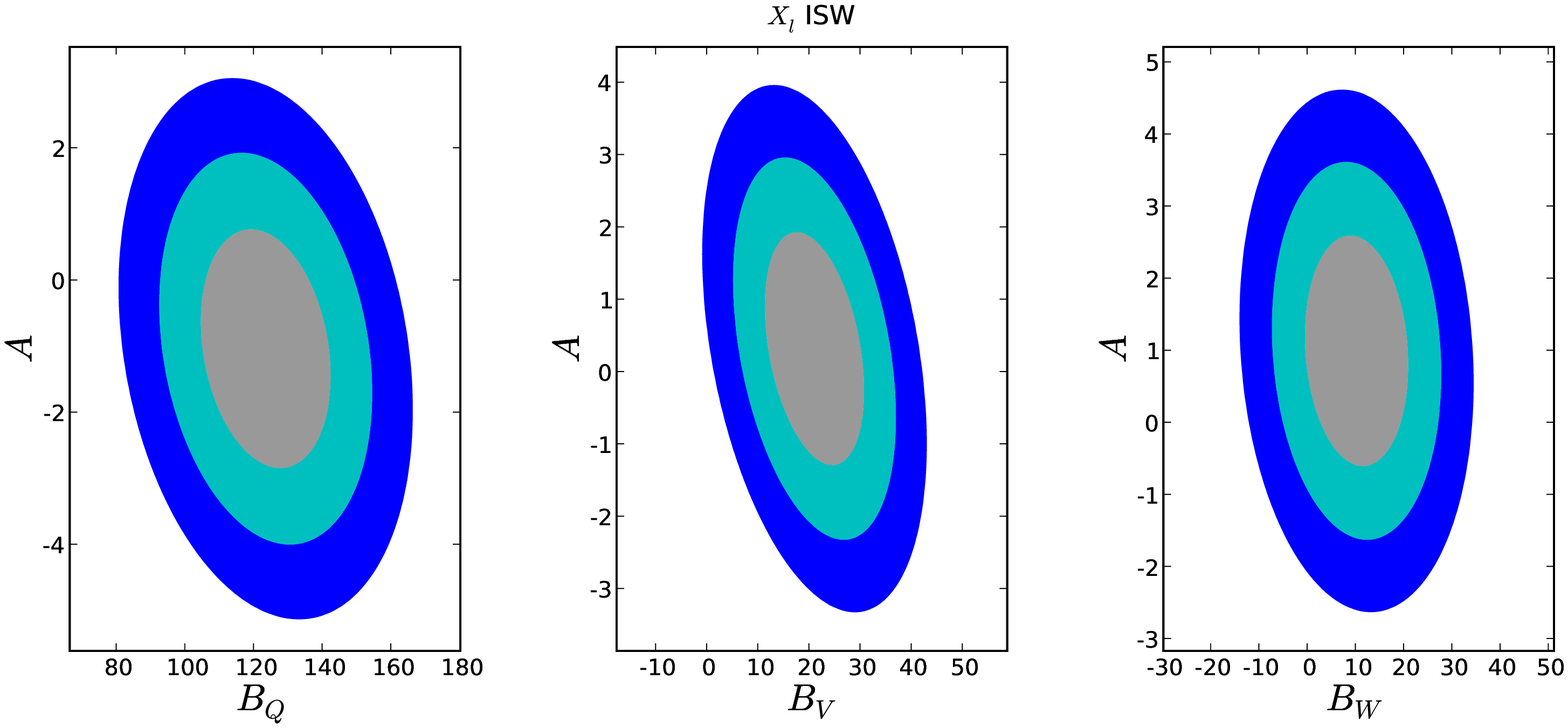,width=0.5\linewidth,clip=} & 
\epsfig{file=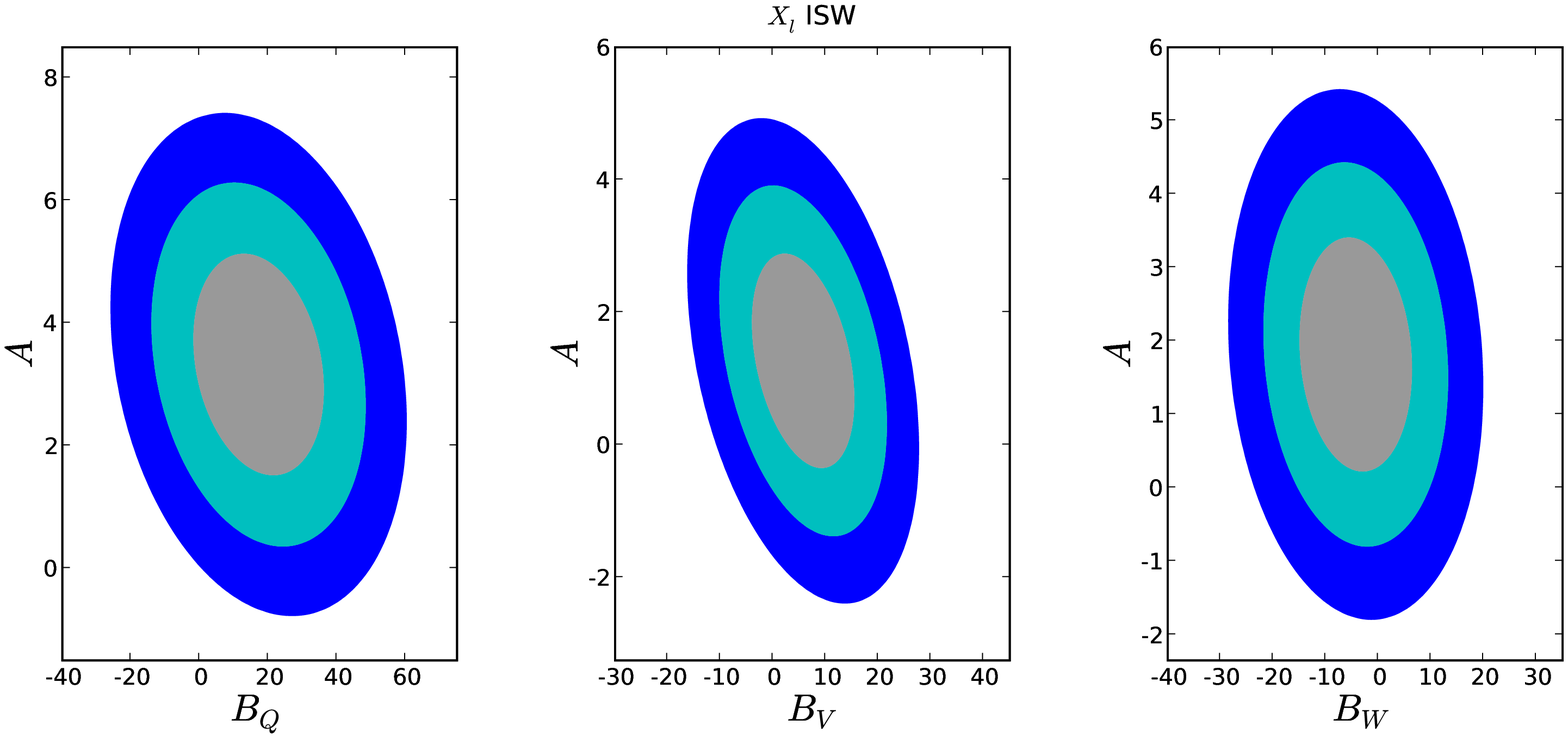,width=0.5\linewidth,clip=} \\
\epsfig{file=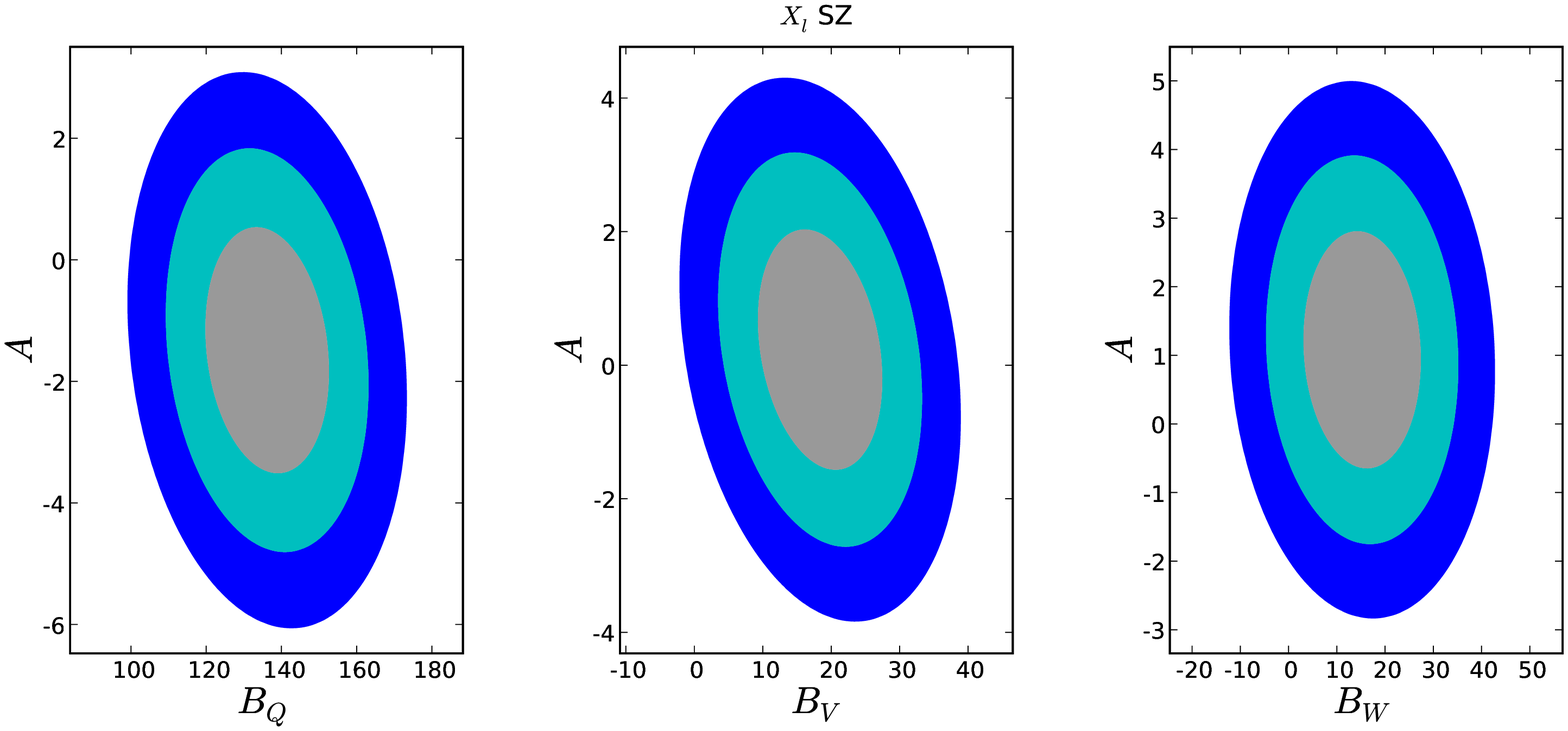,width=0.5\linewidth,clip=} &
\epsfig{file=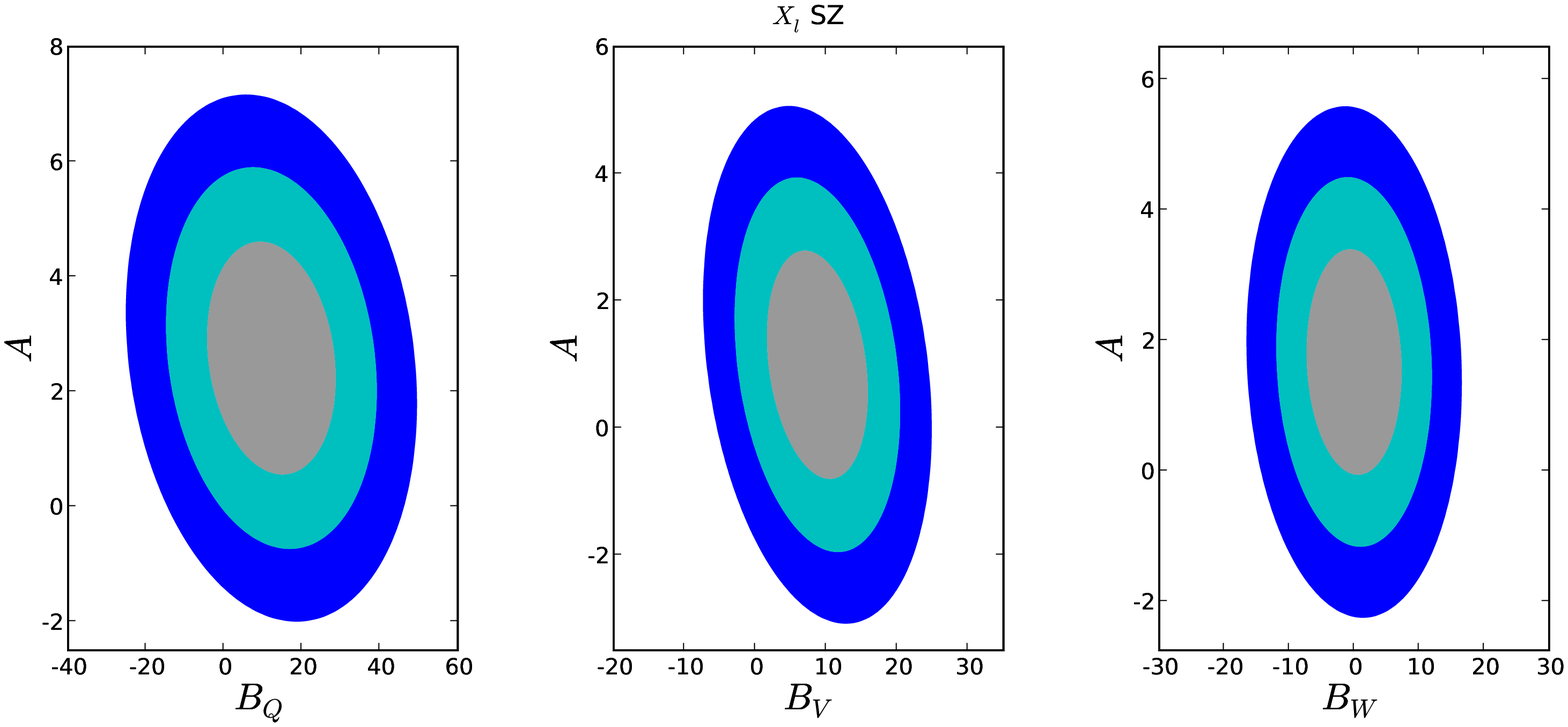,width=0.5\linewidth,clip=}\\
\epsfig{file=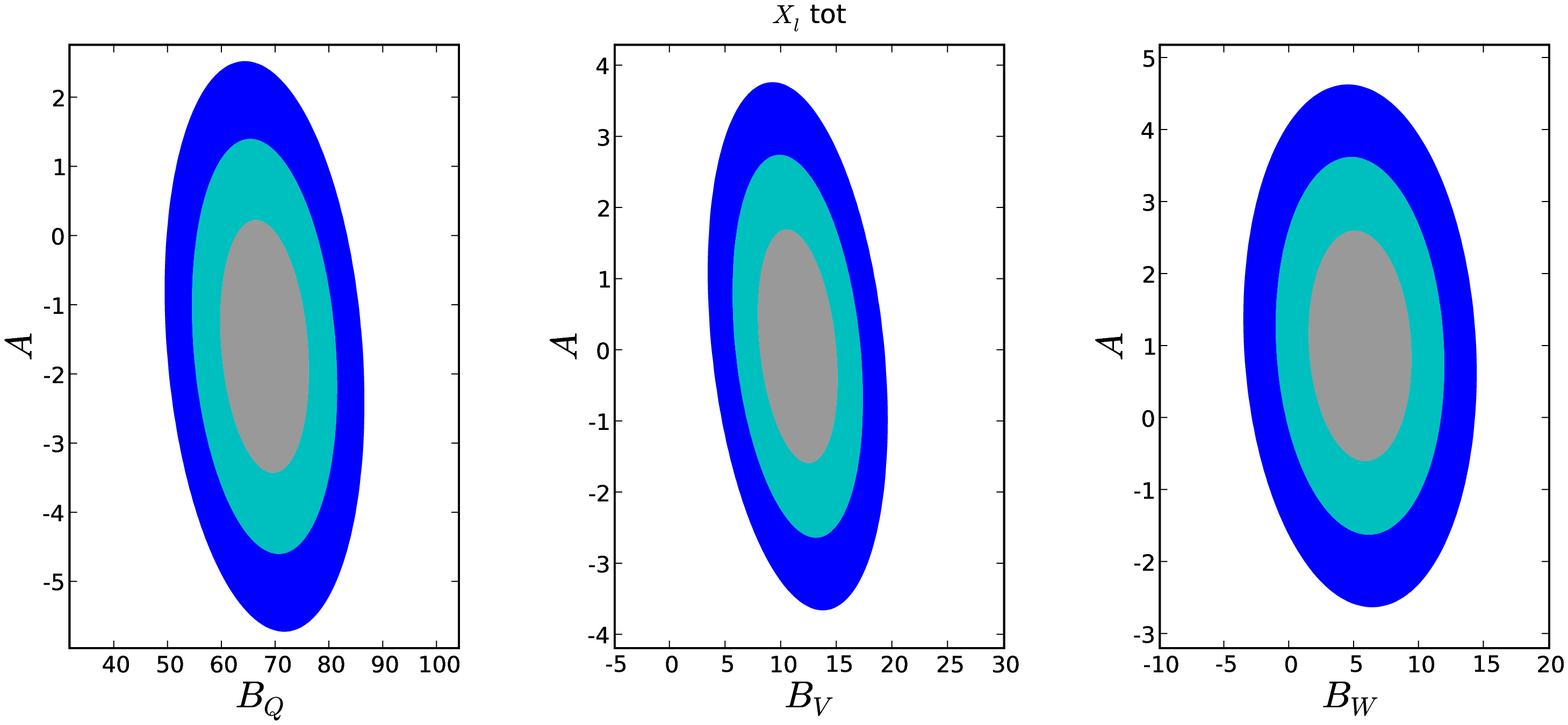,width=0.5\linewidth,clip=} &
\epsfig{file=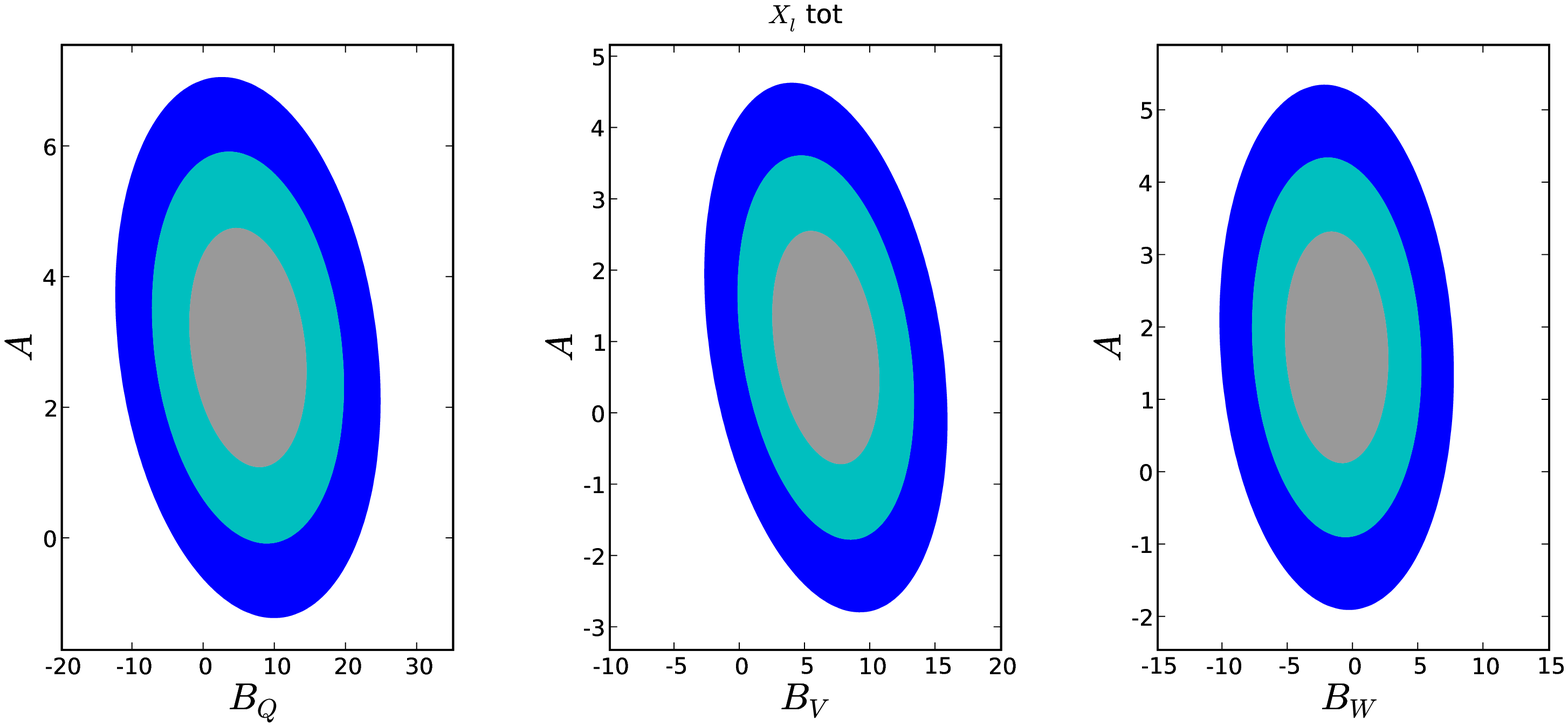,width=0.5\linewidth,clip=}\\
\ \ (a) & \ \ (b)
\end{tabular}
\centering
\caption{\label{degeneracy} 2-dimensional countour plots showing the degeneracies at $68 \%$, $95 \%$ and $99.7 \%$ 
confidence levels between the best fit parameters from raw (left panel (a)) and clean (right panel (b)) map data analysis
for the ISW (top), SZ (middle) and joint ISW+SZ (bottom) cases.}
\end{figure*}

In Figure \ref{degeneracy} we report the two dimensional countour plots showing degeneracies between our best fit parameters for raw and clean maps from the WMAP team used for the data analysis.

\begin{table*}[!htb]
\begin{center}
\begin{tabular}{@{\ \ }c @{\ \ }|@{\ \ }c@{\ \ }|@{\ \ }c@{\ \ }|@{\ \ }c@{\ \ }|@{\ \ }c@{\ \ }|@{\ \ }c@{\ \ }|@{\ \ }c@{\ \ }|@{\ \ }c @{\ \ }}
Frequency & Data & $A$ & $B_Q$ & $B_V$ & $B_W$ & $f_{NL}$ & $\chi^2/dof$ \\
\hline
Q & Raw & $0.39 \pm 1.99$ & $69.8 \pm 5.6$ & & & $-95 \pm 76$ & $1.68$\\
V & & $-0.51 \pm 1.78$ & &  $11.3 \pm 2.4$ & & $31 \pm 70$ & $0.92$ \\
W & & $2.18 \pm 1.78$  & & & $5.6 \pm 2.6$ &  $-60 \pm 73$ & $0.56$\\
Q+W+V & & $1.58 \pm 1.46$ & $68.6 \pm 5.2$ &$12.2 \pm 2.1$& $5.8 \pm 2.5$& $-70 \pm 56$& $0.60$ \\
W+V & & $0.75 \pm 1.56$ & & $11.9 \pm 2.2$ & $5.1 \pm 2.5 $ &$ -16 \pm 62$& $0.85$\\
\hline
Q & Clean & $2.02 \pm 1.99$ & $5.3 \pm 5.6$ & & & $43 \pm 76$ & $0.58$\\
V & & $-0.03 \pm 1.78$ & &  $4.2 \pm 2.4$ & & $48 \pm 70$ & $0.73$ \\
W & & $3.04 \pm 1.78$  & & & $-1.1 \pm 2.6$ & $-67 \pm 73$ & $0.42$\\
Q+W+V & & $1.59 \pm 1.46$ & $8.9 \pm 5.2$ &$4.1 \pm 2.2$& $-1.4 \pm 2.5$& $0 \pm 56$& $0.39$ \\
W+V & & $1.47 \pm 1.56$ & & $4.9 \pm 2.3$ & $-1.6 \pm 2.5$ & $-13 \pm 62$ & $0.80$\\
\end{tabular}
\caption{Amplitude parameters estimation using WMAP raw and clean maps for $X_l$ total and including an extra parameter related to primordial
non-Gaussianity.}
\label{par3}
\end{center}
\end{table*}

\section{Conclusions}

We measure the skewness power spectrum of the CMB anisotropies optimized for a detection of the secondary bispectrum generated by
the correlation of the CMB lensing potential with integrated Sachs-Wolfe effect and the Sunyaev-Zel'dovich effect. The covariance
of our measurements are generated by Monte-Carlo simulations of Gaussian CMB fields with noise properties consistent with WMAP. 

When interpreting multi-frequency measurements we also take into
account the confusion resulting from the unresolved radio point sources. We analyze Q, V and W-band WMAP 5-year raw and foreground
cleaned maps using the KQ75 mask out to $l_{\rm max}=600$. 

While with the raw maps we find no evidence for a non-zero
non-Gaussian signal from the lensing-secondary correlation in any of the three bands, we find 2$\sigma$ and 3$\sigma$ evidence for  a non-zero amplitude of both
the lensing-ISW and lensing-SZ signals in the foreground cleaned Q-band map provided by the WMAP team, respectively. 
The point source amplitude at the bispectrum level measured with this skewness power spectrum is consistent with
previous measurements using the optimized skewness of the WMAP team's analysis
and a different form of the skewness power spectrum optimized for point sources.   

Finally, as the focus is on secondary non-Gaussianity, the estimator is not optimised to detect primordial non-Gaussian signals, and we find a limit on local type of $f_{\rm NL} = -13 \pm 62$ from cleaned V+W maps.

\end{document}